\documentclass{article}
\usepackage{spconf,amsmath,graphicx}
\usepackage{hyperref}
\usepackage{xcolor}
\usepackage{adjustbox}
\usepackage{multirow}
\usepackage{bbding}
\usepackage{cite}
\usepackage{booktabs}
\usepackage{amssymb}
\usepackage{enumitem}
\newcommand{\eg}{\textit{e.g\@.}}

\title{BERTector: An Intrusion Detection Framework Constructed via Joint-dataset Learning Based on Language Model}

\name{\ninept
\parbox{\linewidth}{\centering Haoyang Hu$^{*}$, Xun Huang$^{*}$, Chenyu Wu, Shiwen Liu, Zhichao Lian, Shuangquan Zhang$^{\dagger}$ \\
\ninept School of Cyber Science and Engineering, Nanjing University of Science and Technology, Nanjing, China \thanks{$^{*}$ Equal contribution, $^{\dagger}$ Corresponding author: \href{zhangsq@njust.edu.cn}{zhangsq@njust.edu.cn}.}\thanks{Our code is availiable at \href{https://github.com/ALIENHHY/BERTector}{this repository}.}}}

\address{}

\begin{document}
\maketitle
\begin{abstract}
Intrusion detection systems (IDS) are widely used to maintain the stability of network environments, but still face restrictions in generalizability due to the heterogeneity of network traffics. In this work, we propose \textit{BERTector}, a new framework of joint-dataset learning for IDS based on BERT. \textit{BERTector} integrates three key components: NSS-Tokenizer for traffic-aware semantic tokenization, supervised fine-tuning with a hybrid dataset, and low-rank adaptation for efficient fine-tuning. Experiments show that \textit{BERTector} achieves state-of-the-art detection accuracy, strong generalizability, and excellent robustness. \textit{BERTector} achieves the highest accuracy of $99.28\%$ on NSL-KDD and reaches the average $80\%$ detection success rate against four perturbations. These results establish a unified and efficient solution for modern IDS in complex and dynamic network environments.
\end{abstract}
\begin{keywords}
IDS, BERT, Joint-dataset Learning
\end{keywords}
%

%%%%%%%%%%%%%%%%%%%%%%%%%%%%%%%%%%%%%%%%%%%%%%%
\section{Introduction}
%%%%%%%%%%%%%%%%%%%%%%%%%%%%%%%%%%%%%%%%%%%%%%%
\label{SOTA}

\begin{figure}[!t]
  \label{method}
  \centering
  \includegraphics[width=\linewidth]{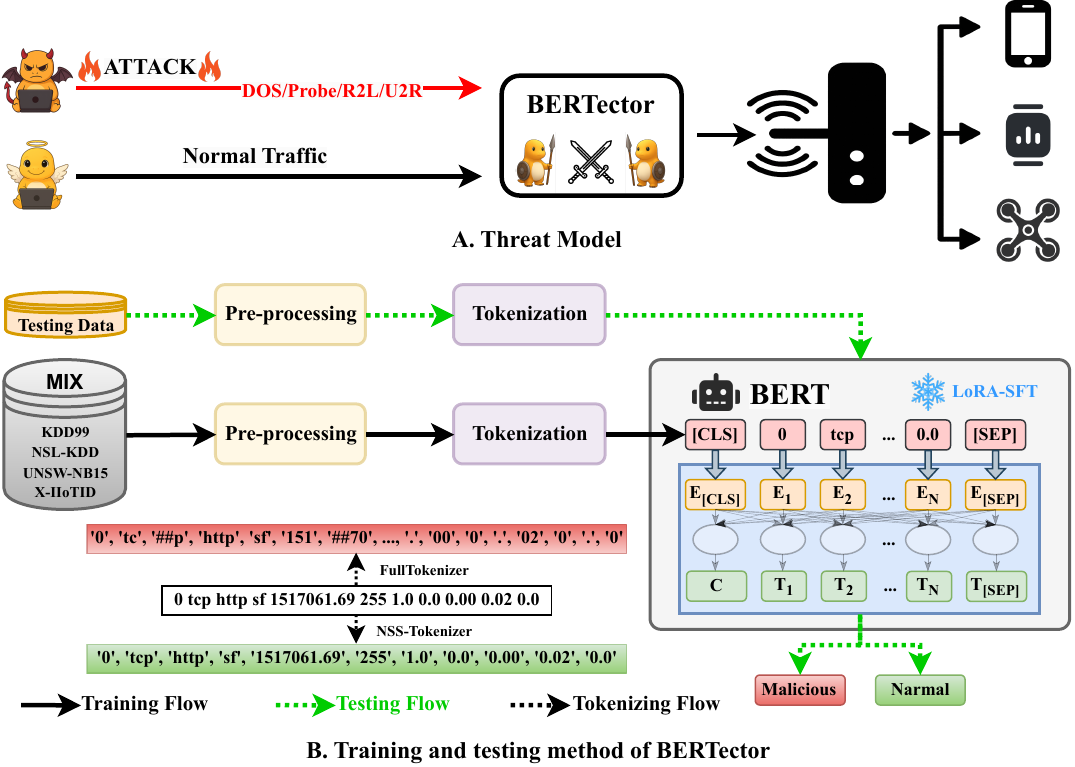}
  \caption{Thread model and the overview of \textit{BERTector}.}
\end{figure}

With the diversification and complexity of network attack methods, key technologies of intrusion detection system (IDS) have gradually transitioned from classic rule-based matching and statistical analysis~\cite{namjoshi2010robust,van2006high} to intelligent detections driven by machine learning (ML) or deep learning (DL)~\cite{lecun2015deep} (see Figure~\ref{method}-A). In past research, many IDS methods have been proposed~\cite{fu2023detecting,yang2025lightweight,zhang2024efficient,fu2024detecting,tsourdinis2024ai,han2022tow,zha2025nids,channappayya2023augmented,yang2024recda,ding2024divide}. Specifically, BIIR~\cite{heidari2023secure} trains RBFNN with Q-learning strategy to further improve the model's performance. AOC-IDS~\cite{zhang2024aoc} uses autoencoders as well as contrastive learning to achieve autonomous online intrusion detection; IG~\cite{pai2024interpretable} introduces a pattern recognition-based approach that can provide efficient detection results even with low training data ratios while enhancing the interpretability of the model; and FLASH~\cite{rehman2024flash} uses an embedder to convert traffic into vectors and then input into XGBoost for classification.

However, these methods have significantly mitigated the risks, they still face serious generalization and robustness issues in attack scenarios with highly diverse attacks and protocol types. Specifically, models trained on a single dataset lack generalizability and are difficult to directly transfer to new scenarios. They require re-training to adapt the new scenarios. Through systematic research, we discover that language models (LMs), with their ability to capture long-sequence semantic associations and the potential for cross-domain feature transfer, can precisely solve the above limitations~\cite{li2025idsagent}. By modeling the global dependencies of traffic sequences, LMs can capture complex attack patterns and potential threats. However, directly applying LMs to IDS faces challenges: (1) network traffic is not natural language, thus its structural and protocol characteristics make the segmentation difficult; (2) standard conversational models with strong capabilities but large parameter scales, making deployment and fine-tuning expensive; (3) attackers may attack with multiple methods and targets, which poses a challenge to adapt the cross-domain detection; (4) adversarial perturbations may be used to evade detection systems.

\noindent\textbf{Our Contributions.} To  address these issues, we propose a BERT-based~\cite{devlin2019bert} scalable \textit{BERTector} framework, which mainly includes: (1) a tokenizer NSS-Tokenizer dedicated for network traffic tokenization; (2) a BERT-based classifier, which contains a small number of parameters but maintains excellent language understanding capability; (3) a multi-source joint training set MIX that integrates numerous open-source datasets to improve the generalizability and robustness; and (4) we fine-tune BERT on MIX and combine it with low-rank adaptation (LoRA)~\cite{hu2022lora} technique to reduce time and computing resource costs (see Figure~\ref{method}-B and \S\ref{sec:Methodology}).

To validate the effectiveness of BERTector, we conduct a comprehensive evaluation. The comparison with baseline on NSL-KDD shows that \textit{BERTector} achieves the highest accuracy of 0.9928, recall of 0.9989, and F1-score of 0.9934, while the precision is also close to the optimal (see \S\ref{Comparison with Baselines}). In the generalizability test, BERTector achieves the accuracy of 0.9887 on KDD99, 0.9610 on UNSW-NB15, and 0.9987 on X-IIoTID. Our method obtains improvements of 6.27\%, 36.20\%, and 73.75\%, compared to the same model trained on the single dataset NSL-KDD. Finally, four perturbations are applied to NSL-KDD, \textit{BERTector} still achieves the accuracy of 0.9374, 0.7678, 0.7336, and 0.7407, respectively, significantly outperforming the baselines. 
%In terms of the ablation study (see \S\ref{Ablation Study}), experimental results show that the components complement each other well. 
In general, our method performs excellently in terms of detection effectiveness, generalizability, and robustness, provides a new approach for IDS in such scenarios insensitive to latency or cost.

%%%%%%%%%%%%%%%%%%%%%%%%%%%%%%%%%%%%%%%%%%%%%%%
\section{Methodology}
%%%%%%%%%%%%%%%%%%%%%%%%%%%%%%%%%%%%%%%%%%%%%%%

\label{sec:Methodology}

\subsection{NSS-Tokenizer}

\label{sec:NSS-Tokenizer}

\begin{table*}[!t]
\caption{Comparison Between NSS-Tokenizer and FullTokenizer on the test set with 10,000 records.}
\label{COMPARISON BETWEEN NSS-TOKENIZER AND BERT-TOKENIZER}
\centering
\begin{adjustbox}{width=0.8\textwidth,center}
\begin{tabular}{cccccccc}
\hline\hline

\multirow{2}{*}{Dataset}& \multirow{2}{*}{Perturbation} & \multicolumn{2}{c}{MAX\_Length (tokens)} & \multicolumn{2}{c}{Pred. Time (s)} & \multicolumn{2}{c}{Tokenize Time (s)} \\
&  & FullTokenizer & \textbf{NSS-Tokenizer} & FullTokenizer & \textbf{NSS-Tokenizer} & FullTokenizer & \textbf{NSS-Tokenizer} \\
 
\hline\hline
NSL-KDD & - & 123 & 41 {\scriptsize\textcolor{red}{↓82}} & 25 & 14 {\scriptsize\textcolor{red}{↓11}} & 4.3274 & 0.0160 {\scriptsize\textcolor{red}{↓4.3114}} \\
KDD99 & - & 111 & 38 {\scriptsize\textcolor{red}{↓73}} & 21 & 13 {\scriptsize\textcolor{red}{↓8}} & 3.9819 & 0.0120 {\scriptsize\textcolor{red}{↓3.9699}} \\
UNSW-NB15 & - & 163 & 43 {\scriptsize\textcolor{red}{↓120}} & 34 & 16 {\scriptsize\textcolor{red}{↓18}} & 6.6414 & 0.0120 {\scriptsize\textcolor{red}{↓6.6294}} \\
X-IIoTID & - & 331 & 65 {\scriptsize\textcolor{red}{↓266}} & 74 & 26 {\scriptsize\textcolor{red}{↓48}} & 11.5066 & 0.0120 {\scriptsize\textcolor{red}{↓11.4946}} \\

\hline

\multirow{4}{*}{NSL-KDD}
& Poission & 129 & 41 {\scriptsize\textcolor{red}{↓88}} & 28 & 15 {\scriptsize\textcolor{red}{↓13}} & 4.8898 & 0.0208 {\scriptsize\textcolor{red}{↓4.8690}} \\ 
& Uniform & 487 & 41 {\scriptsize\textcolor{red}{↓446}} & 118 & 30 {\scriptsize\textcolor{red}{↓88}} & 20.5115 & 0.0120 {\scriptsize\textcolor{red}{↓20.4995}} \\
& Gaussian & 485 & 41 {\scriptsize\textcolor{red}{↓444}} & 117 & 30 {\scriptsize\textcolor{red}{↓87}} & 21.1695 & 0.0240 {\scriptsize\textcolor{red}{↓21.1455}} \\
& Laplace & 485 & 41 {\scriptsize\textcolor{red}{↓444}} & 117 & 30 {\scriptsize\textcolor{red}{↓87}} & 21.1055 & 0.0285 {\scriptsize\textcolor{red}{↓21.0770}} \\
\hline\hline
\end{tabular}
\end{adjustbox}
\end{table*}

The commonly used FullTokenizer in BERT is designed for natural language, thus its tokenization strategy cannot capture the structured protocol features and semantic boundaries of network traffic, which will result in excessive information redundancy, long length of sequences, and inefficient learning. To this end, we propose the NSS-Tokenizer (see Figure~\ref{method}-B), which is specifically designed for traffic flow formats. NSS-Tokenizer uses a dynamic window to control the number of tokens as defined in Eq.\eqref{nss-tokenizer}, where $\text{len}(f)$ represents the length of traffic flows, and $D_{\text{train}}$ denotes training set. The tokenization strategy based on the feature boundary will identify special symbols (\eg, commas, exclamation marks \textit{et al.}) that accurately extract protocol fields and traffic features, reducing the generation noise of invalid tokens, and significantly shorten the input sequence. In addition, NSS-Tokenizer can uniformly tokenize multi-source heterogeneous traffic with different dimensions, keeping the model input consistent, and provide a unified feature expression for subsequent supervised fine-tuning of joint datasets (see Figure~\ref{method}-B). As shown in Table~\ref{COMPARISON BETWEEN NSS-TOKENIZER AND BERT-TOKENIZER}, NSS-Tokenizer performs significantly better than FullTokenizer in terms of token length, model inference latency, and tokenization cost.

\begin{equation}
    \label{nss-tokenizer}
    \text{window} = \min\left( \max\left(\{ \text{len}(f) \} \mid f \in D_{\text{train}} \right), 512 \right)
\end{equation}

\subsection{Joint-dataset construction}

In order to improve the generalizability of \textit{BERTector}, we screen a batch of publicly available traffic datasets from actual network security threats. For the feature fields of different datasets, we use a special symbol that does not appear in the traffic data to separate them, ensuring the integrity of data structure and feature information. In this process, there is no need to worry about the inconsistency of the number of features in different datasets because LM can treat it as a continuous text flow for modeling. With the help of the special symbol segmentation mechanism of NSS-Tokenizer (see \S\ref{sec:NSS-Tokenizer}), each value of features is divided into an independent token, and the traffic will be parsed from the perspective of language modeling, which not only maintains semantic integrity, but also flexibly aligns label information. This design fully utilizes the advantages of LM over ML or DL methods, laying the foundation for building a high-quality joint-dataset and achieving unified training.

\subsection{LoRA Based Supervised Fine-tune}

\noindent \textbf{Supervised Fine-tune (SFT).} To further enhance the perception of model's task and the cross-dataset generalizability, we employ the SFT and combine with label-sensitive cross-entropy loss to detect attack categories on a joint-dataset. During the fine-tuning process, dropout and early stopping were used to improve noise resistance and avoid overfitting. Through joint-dataset training, SFT enables \textit{BERTector} to accurately detect various types of attack in heterogeneous and complex network traffic scenarios, significantly expanding the generalizability and practicality of our mwthod. \\

\noindent \textbf{Low-Rank Adaptation (LoRA).} Although full parameter fine-tuning can maximize the performance of BERT, the high training cost greatly limits its practical application. To this end, we introduce LoRA defined in Eq.\eqref{lora}. For a pre-trained weight matrix $W_0$, we constrain its update by representing the latter with a low-rank decomposition $W_0+\Delta W=W_0+BA$, where $A \in \mathbb{R}^{r \times k}$ and $B \in \mathbb{R}^{d \times r}$ are trainable low-rank matrices, with rank $r \ll \min(k, d)$. During training, only the parameters of $A$ and $B$ are updated, while the original weight matrix $W_0$ remains frozen. In the update term $\Delta W = BA$, both are multiplied by the same input, and their respective output vectors are added coordinately. In the evaluation stage, the combined matrix $W_0 + BA$ is used directly, with no additional computational overhead. By training on multi-source mixed datasets, LoRA can reduce training time while effectively learning the effective features of the joint dataset.

\begin{equation}
    \label{lora}
    h=W_0x+\Delta Wx=W_0x+BAx
\end{equation}

%%%%%%%%%%%%%%%%%%%%%%%%%%%%%%%%%%%%%%%%%%%%%%%
\section{Experiments}
%%%%%%%%%%%%%%%%%%%%%%%%%%%%%%%%%%%%%%%%%%%%%%%

\subsection{Experimental Setup}

\noindent\textbf{Datasets.}
To verify the versatility and cross-domain adaptability, we construct a joint-dataset MIX integrating four commonly-used datasets: (1) KDD-99~\cite{kdd_cup_1999_data_130}, (2) NSL-KDD~\cite{tavallaee2009detailed}, (3) UNSW-NB15~\cite{moustafa2015unsw}, and (4) X-IIoTID~\cite{al2021x}. Each dataset is uniformly converted to net-flow format and semantically tokenized using NSS-Tokenizer. MIX samples 100,000 records from each source dataset and is split into training and validation sets with a ratio of 4:1 to ensure diversity and coverage, while every test set contains 10,000 non-repeat records from each of four datasets for evaluation. \\

\noindent\textbf{Metrics.}
Accuracy, precision, recall, and F1-score are used as evaluation metrics. Accuracy reflects the overall quality of the predictions; precision measures the proportion of correct detections; recall indicates the proportion of successful attacks identified; and the F1-score harmonizes precision and recall to deliver a single, balanced view of performance. \\

\noindent\textbf{Environment.}
All experiments are carried out on a Windows 10 system equipped with a NVIDIA GeForce RTX 4090 GPU (24GB VRAM), and an i9-13900kf CPU (48GB RAM). The learning rate was set to $2\times10^{-5}$, with a batch size of 64, as well as 10 training epochs. L2 regularization was applied and early-stopping was employed to prevent overfitting.

\subsection{Comparison with Baselines}

\label{Comparison with Baselines}

We compare  BERTector against baselines including classic ML models \eg, Random Forest (RF), Decision Tree (DT), and Linear Regression (LR); as well as DL models \eg, DNN, RNN, LSTM, GBM, and XGBoost~\cite{chen2016xgboost}. A fair comparison was ensured by applying appropriate feature engineering and hyperparameter optimization to all models, allowing each method to perform optimally. Furthermore, we directly introduce the experimental results of several SOTA methods (see \S\ref{SOTA}). As shown in Table~\ref{Comparison of BERTector and comparison Models on NSL-KDD}, \textit{BERTector} achieves an accuracy of 0.9928 and an F1-score of 0.9934; while its Precision is slightly lower than XGBoost, this minor gap does not affect the overall performance superiority of \textit{BERTector}, while compared to previous SOTA methods, \textit{BERTector} also achieves a better effectiveness. These results illustrate the exceptional detection capability of our method when processing complex, multi-dimensional network traffic patterns.

\begin{table}[!t]
\centering
\caption{Comparison with baselines on NSL-KDD.}
\label{Comparison of BERTector and comparison Models on NSL-KDD}
\begin{adjustbox}{width=0.45\textwidth,center}
\begin{tabular}{cccccc}
\hline\hline
 & & Accuracy & Precision & Recall & F1-score \\
\hline\hline
 & RF & 0.9498 & 0.9885 & 0.9181 & 0.9520 \\
 & DT & 0.8447 & 0.9293 & 0.7725 & 0.8437 \\
ML & LR & 0.9394 & 0.9412 & 0.9475 & 0.9443 \\
 & GBM & 0.8911 & 0.9835 & 0.8129 & 0.8901 \\
 & XGBoost & 0.9307 & \textbf{0.9935} & 0.8779 & 0.9322 \\
\hline
 & DNN & 0.9912 & 0.9904 & 0.9934 & 0.9919 \\
DL & RNN & 0.9916 & 0.9932 & 0.9913 & 0.9922 \\
 & LSTM & 0.9918 & 0.9915 & 0.9934 & 0.9924 \\
\hline
 & BIIR~\cite{heidari2023secure} & 0.9921 & 0.9930 & 0.9940 & 0.9910 \\
SOTA & AOC-IDS~\cite{zhang2024aoc} & 0.8890 & 0.8599 & 0.9621 & 0.9081 \\
 & IG~\cite{pai2024interpretable} & 0.9389 & 0.9353 & 0.9426 & 0.9402 \\
\hline
Ours & \textbf{BERTector} & \textbf{0.9928} & 0.9880 & \textbf{0.9989} & \textbf{0.9934} \\
\hline\hline
\end{tabular}
\end{adjustbox}
\end{table}

\subsection{Generalizability Evaluation}

To systematically evaluate the generalizability of our method, we jointly fine-tune \textit{BERTector} on the MIX dataset, during which \textit{BERTector} could learn various traffic features rather than adapting to a specific format of dataset. After fine-tuning, we test \textit{BERTector} on each single dataset separately to verify the migration capabilities of our method under different traffic attacks, protocol types, and scenarios. ``BERTector-MIX'' shows the most strongest generalizability among the test sets, especially on KDD99, UNSW-NB15 and X-IIoTID, with accuracy of 0.9887, 0.9610, and 0.9987 respectively, which far outperforms the models trained on a single dataset (see Table~\ref{Cross-datasets Generalization Testing of BERTector.}). In contrast, the ``BERT+SFT+NSS'' without joint training achieves the greatest results on the specific dataset with prior exposure, but its migration capability in other scenarios is limited. The experimental results verify that joint training on the MIX datasets can effectively improve the generalizability of the detecting model and the versatility of application scenarios.

\begin{table}[!t]
\centering
\caption{Generalizability Evaluation of \textit{BERTector}.}
\label{Cross-datasets Generalization Testing of BERTector.}
\begin{adjustbox}{width=0.45\textwidth,center}
\begin{tabular}{ccccc}
\hline\hline
 & NSL-KDD & KDD99 & UNSW-NB15 & X-IIoTID \\
\hline\hline
BERT+SFT & 0.9822 & 0.8496 & 0.1196 & 0.3960   \\
BERT+SFT+LoRA & 0.9157 & 0.5112 & 0.0820 & 0.5174   \\
BERT+SFT+NSS & \textbf{0.9980} & 0.8473 & 0.7744 & 0.4520   \\
\textbf{BERTector}& 0.9928 & 0.9304 & 0.7056 & 0.5748   \\
\hline
\textbf{BERTector-MIX} & 0.9903 & \textbf{0.9887} & \textbf{0.9610} & \textbf{0.9987}   \\
\hline\hline \\
\end{tabular}
\end{adjustbox}
\end{table}

\begin{table}[!t]
\centering
\caption{Robustness Test on NSL-KDD.}
\label{Robustness Test Results on Perturbed NSL-KDD}
\begin{adjustbox}{width=0.45\textwidth,center}
\begin{tabular}{cccccc}
\hline\hline
Perturbation & Models & Accuracy & Precision & Recall & F1-score \\
\hline\hline
& RF        & 0.8026 & 0.8781    & 0.7386 & 0.8023   \\
& DT        & 0.7329 & 0.7447    & 0.7723 & 0.7583   \\
& LR        & 0.8172 & 0.8596    & 0.7924 & 0.8246   \\
& GBM       & 0.8279 & \textbf{0.9621} & 0.7107 & 0.8175   \\
\textbf{Poission} & XGBoost   & 0.8218 & 0.9317 & 0.7246 & 0.8152   \\
& DNN & 0.6582 & 0.6463 & 0.8169 & 0.7217   \\
& RNN       & 0.6617 & 0.6817 & 0.7058 & 0.6935   \\
& LSTM      & 0.6805 & 0.6923 & 0.7399 & 0.7153   \\
\cline{2-6}
& \textbf{BERTector} & \textbf{0.9374} & 0.9209    & \textbf{0.9677} & \textbf{0.9437}   \\

\hline

& RF        & 0.6327 & 0.7014 & 0.5621 & 0.6241   \\
& DT        & 0.6033 & 0.6470 & 0.5913 & 0.6179   \\
& LR        & 0.6067 & 0.6557 & 0.5787 & 0.6148   \\
& GBM       & 0.6107 & 0.7148 & 0.4696 & 0.5668   \\
\textbf{Uniform} & XGBoost   & 0.6277 & 0.7281 & 0.5006 & 0.5932   \\
& DNN & 0.5323 & 0.5544 & 0.7017 & 0.6194   \\
& RNN       & 0.5304 & 0.5669 & 0.5684 & 0.5677   \\
& LSTM      & 0.5304 & 0.5668 & 0.5697 & 0.5682   \\
\cline{2-6}
& \textbf{BERTector} & \textbf{0.7678} & \textbf{0.7463} & \textbf{0.8663} & \textbf{0.8019}   \\

\hline

& RF        & 0.6549 & 0.7329 & 0.5723 & 0.6427   \\
& DT        & 0.6043 & 0.6495 & 0.5874 & 0.6169   \\
& LR        & 0.6433 & 0.6917 & 0.6176 & 0.6526   \\
& GBM       & 0.6465 & \textbf{0.7952} & 0.4690 & 0.5900   \\
\textbf{Gaussian} & XGBoost   & 0.6440 & 0.7663 & 0.4945 & 0.6011   \\
& DNN & 0.5386 & 0.5584 & 0.7140 & 0.6267   \\
& RNN       & 0.5317 & 0.5675 & 0.5741 & 0.5708   \\
& LSTM      & 0.5404 & 0.5751 & 0.5848 & 0.5799   \\
\cline{2-6}
& \textbf{BERTector} & \textbf{0.7336} & 0.7055 & \textbf{0.8733} & \textbf{0.7805}   \\

\hline

& RF        & 0.6534 & 0.7486 & 0.5435 & 0.6298   \\
& DT        & 0.5936 & 0.6385 & 0.5778 & 0.6067   \\
& LR        & 0.6685 & 0.7198 & 0.6366 & 0.6757   \\
& GBM       & 0.6517 & \textbf{0.8341} & 0.4467 & 0.5818   \\
\textbf{Laplace} & XGBoost   & 0.6297 & 0.7658 & 0.4570 & 0.5725   \\
& DNN & 0.5424 & 0.5616 & 0.7131 & 0.6283   \\
& RNN       & 0.5391 & 0.5754 & 0.5736 & 0.5745   \\
& LSTM      & 0.5444 & 0.5791 & 0.5861 & 0.5826   \\
\cline{2-6}
& \textbf{BERTector} & \textbf{0.7407} & 0.7115 & \textbf{0.8779} & \textbf{0.7860}   \\

\hline\hline
\end{tabular}
\end{adjustbox}
\end{table}

\subsection{Robustness Test}

To verify the robustness of \textit{BERTector} under adversarial attacks, we introduce Poisson, Uniform, Gaussian, and Laplace perturbations~\cite{diochnos2018adversarial,erdemir2021adversarial,hendrycks2019benchmarking,li2019certified,nasr2021defeating,moosavi2017universal,zhang2021adversarial} on the NSL-KDD test set. Each method is used to simulate the attacker's numerical interference on the original traffic to avoid being detected by IDS, thus testing the detection stability of our method in such scenarios. We select classic ML and DL methods to make a comparison with \textit{BERTector}. All models are trained on the basis of normal samples, and perturbations are only applied during the test phase to objectively evaluate their robustness. Table~\ref{Robustness Test Results on Perturbed NSL-KDD} shows the experimental results that \textit{BERTector} performs significantly more robustness under all types of perturbation. Under Poisson disturbance, \textit{BERTector} achieves an accuracy of 0.9374 and an F1-score of 0.9437, far exceeding other comparison methods. Even under perturbations with higher interference intensity: Uniform, Gaussian, and Laplace, \textit{BERTector} still maintained accuracy of 0.7678, 0.7336 and 0.7407, with each F1 score higher than 0.78. In contrast, traditional ML and DL methods show significant degradation under these perturbations, especially in strong noise environments such as Uniform and Gaussian, their accuracy typically collapses below 65\%. These results verify the robustness of our method, which can effectively combat various types of adversarial perturbation.

\begin{table}[t!]
\caption{Ablation Study on NSL-KDD.}
\label{ABLATION TEST RESULTS}
\centering
\begin{adjustbox}{width=0.45\textwidth,center}
\begin{tabular}{ccc|ccccc}
\hline\hline
SFT & NSS & LoRA & Time (s) & Accuracy & Precision & Recall & F1-score \\
\hline\hline
\XSolidBrush & \XSolidBrush & \XSolidBrush & - & 0.3095 & 0.1965 & 0.0883 & 0.1219 \\
\CheckmarkBold & \XSolidBrush & \XSolidBrush & 2015 & 0.9822 & 0.9776 & 0.9899 & 0.9837 \\
\CheckmarkBold & \CheckmarkBold & \XSolidBrush & 813 & 0.9980 & 0.9971 & 0.9993 & 0.9982 \\
\CheckmarkBold & \XSolidBrush & \CheckmarkBold & 1530 & 0.9157 & 0.8717 & 0.9904 & 0.9272 \\
\CheckmarkBold & \CheckmarkBold & \CheckmarkBold & 586 & 0.9928 & 0.9880 & 0.9989 & 0.9934 \\
\hline\hline
\end{tabular}
\end{adjustbox}
\end{table}

\subsection{Ablation Study}

\label{Ablation Study}

We conduct ablation study on NSL-KDD. As shown in Table~\ref{ABLATION TEST RESULTS},the base model BERT without any modules achieves the accuracy of 0.3095. The BERT with SFT significantly  obtains accuracy of 0.9822. Further introducing NSS-Tokenizer, ``BERT+NSS+SFT'' achieves state-of-the-art performance across all four metrics: accuracy of 0.9980, precision of 0.9971, recall of 0.9993, and F1-score of 0.9982. Our final solution buils on this foundation by overlaying LoRA, which reduces fine-tuning overhead. Evaluation metrics decrease slightly to 0.9928, 0.9880, 0.9989, and 0.9934, respectively. Fine-tuning time is further reduced from 813 seconds to 586 seconds, achieving a 28\% reduction compared to the optimal combination. Furthermore, removing NSS-Tokenizer achieves a higher recall of 0.9904, the precision dropped significantly to 0.8717, demonstrating that NSS is crucial for suppressing false positives. In general, the final solution ``BERT+NSS+SFT+LoRA'' significantly reduces the time and computation costs with a manageable loss of accuracy, providing a better trade-off.

%%%%%%%%%%%%%%%%%%%%%%%%%%%%%%%%%%%%%%%%%%%%%%%
\section{Conclusion}
%%%%%%%%%%%%%%%%%%%%%%%%%%%%%%%%%%%%%%%%%%%%%%%

This paper highlights the limitations of IDS in complex traffic environments. Specifically, we propose \textit{BERTector}, which is uniformly fine-tuned on a composite corpus MIX covering four commonly-used datasets, while integrating NSS-Tokenizer and LoRA adapter to improve the effectiveness. Extensive experiments demonstrate that this design not only eliminates the need for scenario-specific re-training but also achieves stable cross-domain detection generalizability and robustness, as well as achieving a favorable efficiency-accuracy trade-off. Above all, our research provides a practical and scalable solution for multi-source heterogeneous network. Although it has higher cost and latency than classic ML or DL methods, it still provides an idea to the design of detection devices in cost and latency insensitive dynamic scenarios.

%%%%%%%%%%%%%%%%%%%%%%%%%%%%%%%%%%%%%%%%%%%%%%%
\section{Acknowledgment}
%%%%%%%%%%%%%%%%%%%%%%%%%%%%%%%%%%%%%%%%%%%%%%%

This research is supported by the National Natural Science Foundation of China (No. 62302218), Qing Lan Project, Key R\&D Program of Jiangsu (BE2022081).

\vfill\pagebreak
\ninept
\bibliographystyle{IEEEbib}
\bibliography{BERTector}

\begin{thebibliography}{10}

\bibitem{namjoshi2010robust}
Kedar Namjoshi and Girija Narlikar,
\newblock ``Robust and fast pattern matching for intrusion detection,''
\newblock in {\em Proc. of INFOCOM}, 2010.

\bibitem{van2006high}
Jan Van~Lunteren,
\newblock ``High-performance pattern-matching for intrusion detection,''
\newblock in {\em Proc. of INFOCOM}, 2006.

\bibitem{lecun2015deep}
Yann LeCun, Yoshua Bengio, and Geoffrey Hinton,
\newblock ``Deep learning,''
\newblock {\em Nature}, 2015.

\bibitem{fu2023detecting}
Chuanpu Fu, Qi~Li, and Ke~Xu,
\newblock ``Detecting unknown encrypted malicious traffic in real time via flow interaction graph analysis,''
\newblock in {\em Proc. of NDSS}, 2023.

\bibitem{yang2025lightweight}
Xueji Yang, Fei Tong, Fang Jiang, and Guang Cheng,
\newblock ``A lightweight and dynamic open-set intrusion detection for industrial internet of things,''
\newblock {\em IEEE Transactions on Information Forensics and Security}, 2025.

\bibitem{zhang2024efficient}
Linxi Zhang, Xuke Yan, and Di~Ma,
\newblock ``Efficient and effective in-vehicle intrusion detection system using binarized convolutional neural network,''
\newblock in {\em Proc. of INFOCOM}, 2024.

\bibitem{fu2024detecting}
Chuanpu Fu, Qi~Li, Meng Shen, and Ke~Xu,
\newblock ``Detecting tunneled flooding traffic via deep semantic analysis of packet length patterns,''
\newblock in {\em Proc. of CCS}, 2024.

\bibitem{tsourdinis2024ai}
Theodoros Tsourdinis, Nikos Makris, Thanasis Korakis, and Serge Fdida,
\newblock ``Ai-driven network intrusion detection and resource allocation in real-world o-ran 5g networks,''
\newblock in {\em Proc. of MobiCom}, 2024.

\bibitem{han2022tow}
Mee~Lan Han, Byung~Il Kwak, and Huy~Kang Kim,
\newblock ``Tow-ids: intrusion detection system based on three overlapped wavelets for automotive ethernet,''
\newblock {\em IEEE Transactions on Information Forensics and Security}, 2022.

\bibitem{zha2025nids}
Chao Zha, Zhiyu Wang, Yifei Fan, Bing Bai, Yinjie Zhang, Sainan Shi, and Ruyun Zhang,
\newblock ``A-nids: adaptive network intrusion detection system based on clustering and stacked ctgan,''
\newblock {\em IEEE Transactions on Information Forensics and Security}, 2025.

\bibitem{channappayya2023augmented}
Sumohana Channappayya, Bheemarjuna~Reddy Tamma, et~al.,
\newblock ``Augmented memory replay-based continual learning approaches for network intrusion detection,''
\newblock {\em Proc. of NeurIPS}, 2023.

\bibitem{yang2024recda}
Shuo Yang, Xinran Zheng, Jinze Li, Jinfeng Xu, Xingjun Wang, and Edith~CH Ngai,
\newblock ``Recda: Concept drift adaptation with representation enhancement for network intrusion detection,''
\newblock in {\em Proc. of SIGKDD}, 2024.

\bibitem{ding2024divide}
Hua Ding, Lixing Chen, Shenghong Li, Yang Bai, Pan Zhou, and Zhe Qu,
\newblock ``Divide, conquer, and coalesce: Meta parallel graph neural network for iot intrusion detection at scale,''
\newblock in {\em Proc. of WWW}, 2024.

\bibitem{heidari2023secure}
Arash Heidari, Nima~Jafari Navimipour, and Mehmet Unal,
\newblock ``A secure intrusion detection platform using blockchain and radial basis function neural networks for internet of drones,''
\newblock {\em IEEE Internet of Things Journal}, 2023.

\bibitem{zhang2024aoc}
Xinchen Zhang, Running Zhao, Zhihan Jiang, Zhicong Sun, Yulong Ding, Edith~CH Ngai, and Shuang-Hua Yang,
\newblock ``Aoc-ids: Autonomous online framework with contrastive learning for intrusion detection,''
\newblock in {\em Proc. of INFOCOM}, 2024.

\bibitem{pai2024interpretable}
Hao-Ting Pai, Yu-Hsuan Kang, and Wen-Cheng Chung,
\newblock ``An interpretable generalization mechanism for accurately detecting anomaly and identifying networking intrusion techniques,''
\newblock {\em IEEE Transactions on Information Forensics and Security}, 2024.

\bibitem{rehman2024flash}
Mati~Ur Rehman, Hadi Ahmadi, and Wajih~Ul Hassan,
\newblock ``Flash: A comprehensive approach to intrusion detection via provenance graph representation learning,''
\newblock in {\em Proc. of IEEE S\&P}, 2024.

\bibitem{li2025idsagent}
Yanjie Li, Zhen Xiang, Nathaniel~D. Bastian, Dawn Song, and Bo~Li,
\newblock ``Ids-agent: An llm agent for explainable intrusion detection in iot networks,''
\newblock in {\em Proc. of NeurIPS Workshop}, 2024.

\bibitem{devlin2019bert}
Jacob Devlin, Ming-Wei Chang, Kenton Lee, and Kristina Toutanova,
\newblock ``Bert: Pre-training of deep bidirectional transformers for language understanding,''
\newblock in {\em Proc. of ACL}, 2019.

\bibitem{hu2022lora}
Edward~J Hu, Yelong Shen, Phillip Wallis, Zeyuan Allen-Zhu, Yuanzhi Li, Shean Wang, Lu~Wang, Weizhu Chen, et~al.,
\newblock ``Lora: Low-rank adaptation of large language models.,''
\newblock in {\em Proc. of ICLR}, 2022.

\bibitem{kdd_cup_1999_data_130}
Salvatore~J. Stolfo, Wei Fan, Wenke Lee, Andreas~L. Prodromidis, and Philip~K. Chan,
\newblock ``{KDD Cup 1999 Data},'' UCI Machine Learning Repository, 1999,
\newblock {DOI}: https://doi.org/10.24432/C51C7N.

\bibitem{tavallaee2009detailed}
Mahbod Tavallaee, Ebrahim Bagheri, Wei Lu, and Ali~A Ghorbani,
\newblock ``A detailed analysis of the kdd cup 99 data set,''
\newblock in {\em Proc. of IEEE CISDA}, 2009.

\bibitem{moustafa2015unsw}
Nour Moustafa and Jill Slay,
\newblock ``Unsw-nb15: a comprehensive data set for network intrusion detection systems (unsw-nb15 network data set),''
\newblock in {\em Proc. of IEEE MilCIS}, 2015.

\bibitem{al2021x}
Muna Al-Hawawreh, Elena Sitnikova, and Neda Aboutorab,
\newblock ``X-iiotid: A connectivity-agnostic and device-agnostic intrusion data set for industrial internet of things,''
\newblock {\em IEEE Internet of Things Journal}, 2021.

\bibitem{chen2016xgboost}
Tianqi Chen and Carlos Guestrin,
\newblock ``Xgboost: A scalable tree boosting system,''
\newblock in {\em Proc. of SIGKDD}, 2016.

\bibitem{diochnos2018adversarial}
Dimitrios Diochnos, Saeed Mahloujifar, and Mohammad Mahmoody,
\newblock ``Adversarial risk and robustness: General definitions and implications for the uniform distribution,''
\newblock in {\em Proc. of NeurIPS}, 2018.

\bibitem{erdemir2021adversarial}
Ecenaz Erdemir, Jeffrey Bickford, Luca Melis, and Sergul Aydore,
\newblock ``Adversarial robustness with non-uniform perturbations,''
\newblock in {\em Proc. of NeurIPS}, 2021.

\bibitem{hendrycks2019benchmarking}
Dan Hendrycks and Thomas Dietterich,
\newblock ``Benchmarking neural network robustness to common corruptions and perturbations,''
\newblock in {\em Proc. of ICLR}, 2019.

\bibitem{li2019certified}
Bai Li, Changyou Chen, Wenlin Wang, and Lawrence Carin,
\newblock ``Certified adversarial robustness with additive noise,''
\newblock in {\em Proc. of NeurIPS}, 2019.

\bibitem{nasr2021defeating}
Milad Nasr, Alireza Bahramali, and Amir Houmansadr,
\newblock ``Defeating dnn-based traffic analysis systems in real-time with blind adversarial perturbations,''
\newblock in {\em Proc. of USENIX Security}, 2021.

\bibitem{moosavi2017universal}
Seyed-Mohsen Moosavi-Dezfooli, Alhussein Fawzi, Omar Fawzi, and Pascal Frossard,
\newblock ``Universal adversarial perturbations,''
\newblock in {\em Proc. of CVPR}, 2017.

\bibitem{zhang2021adversarial}
Xingwei Zhang, Xiaolong Zheng, and Wenji Mao,
\newblock ``Adversarial perturbation defense on deep neural networks,''
\newblock {\em ACM Computing Surveys (CSUR)}, 2021.

\end{thebibliography}

\end{document}